\documentclass[aps,prl,showpacs,twocolumn]{revtex4}

\usepackage{graphicx,bbm}

\def\tr{\,{\rm tr}\,}

\def\ave#1{\langle #1 \rangle}
\def\ii{{\rm i}}

\def\tit#1{}
\def\etal#1{ {\em et al.}}

\begin{document}

\title{Long range order in non-equilibrium interacting quantum spin chains}

\author{Toma\v z Prosen$^{1,2}$ and Marko \v Znidari\v c$^1$}
\affiliation{$^1$ Department of Physics, Faculty of Mathematics and Physics,
  University of Ljubljana, Ljubljana, Slovenia \\
$^2$ Department of Physics and Astronomy, University of Potsdam, Potsdam, Germany}

\date{\today}

\begin{abstract}
We conjecture that {\em non-equilibrium boundary conditions} generically trigger {\em  long range order} in non-equilibrium steady states of locally interacting quantum chains. Our result is based on large scale density matrix renormalization group simulations of several models of quantum spin $1/2$ chains which are driven far from equilibrium by coupling to a pair of unequal Lindblad reservoirs attached locally to the ends of the chain.
In particular, we find a phase transition from exponentially decaying to long range spin-spin correlations in integrable Heisenberg XXZ chain by changing the anisotropy parameter. Long range order also typically emerges after breaking the integrability of the model.
\end{abstract}

\pacs{05.70.Ln, 05.30.Rt, 03.65.Yz, 75.10.Pq}
 

\maketitle

Non-equilibrium physics of low dimensional locally interacting quantum many-body systems is becoming 
an increasingly hot topic, in particular due to enormous recent progress in experimental techniques with
ultra-cold atomic gases and optical lattices \cite{bloch}, and arguable connection to important open challenges in condensed matter physics such as high-temperature superconductivity and solid-state quantum computation.
While thermalization in isolated many-body systems seems to be well understood \cite{rigol}, and near
equilibrium physics is essentially given by linear response formalism, far from equilibrium even some very basic concepts remain unclear. 

One of the key questions that we address in this Letter is: Driving a large one-dimensional locally interacting quantum system out of equilibrium by coupling to two unequal reservoirs at the system's ends,
will the {\em non-equilibrium steady state} (NESS) in the bulk be well described by the notion of {\em local equilibrium}? If yes, then the properties of NESS are essentially given by local Gibbs state with a well defined local temperature (and perhaps local chemical potential) field(s). If not, then the local temperature has no meaning, and the physics of far from equilibrium system is qualitatively different than in equilibrium. While Mermin-Wagner's theorem \cite{mermin} prohibits certain types of order at finite-temperature equilibrium, 
(arbitrarily weak) non-equilibrium driving at the boundary of the system breaks not only the translational invariance but typically also the symmetries of the Hamiltonian and opens the possibility of {\em long range order} (LRO).
As has been shown for an open XY spin $1/2$ chain, LRO may set in abruptly by varying the system's parameters through a {\em non-equilibrium quantum phase transition} \cite{XY}. 

Nevertheless, since the XY chain is equivalent to a quasi-free fermionic model, one might argue that LRO might be a non-generic effect which would typically not survive switching on the interaction, or even more so, breaking the integrability of the model. In this Letter we show that this is {\em not} the case by performing large scale density-matrix-renormalization-group (DMRG) simulations of NESS for integrable and non-integrable versions of anisotropic locally interacting Heisenberg (XXZ) spin $1/2$ chain.  By changing the anisotropy parameter we find a clear evidence of non-equilibrium quantum phase transition from exponentially decaying spin-spin correlations to LRO where spin-spin correlations saturate and do not depend on the system size.  Perhaps even more strikingly, we find that LRO is typical for non-integrable deformations of the model. Note that the physics of NESS can be very different from the much studied quench dynamics where one can observe at most a transient non-equilibrium phenomena.

The model we study is an anisotropic nearest-neighbor interacting spin 1/2 Heisenberg XXZ chain
\begin{equation}
H=\sum_{j=1}^{n-1} (\sigma_j^{\rm x} \sigma_{j+1}^{\rm x} +\sigma_j^{\rm y} \sigma_{j+1}^{\rm y}+ \Delta \sigma_j^{\rm z} \sigma_{j+1}^{\rm z}).
\label{eq:H}  
\end{equation}
The Hamiltonian (\ref{eq:H}) is an important model for several reasons. It describes various spin-chain materials~\cite{reviews} like cuprates and as a particularly promising experimental technique, it can be realized with cold atoms in an optical lattice~\cite{Trotzky:08}. From the theoretical side, (\ref{eq:H}) is perhaps the simplest model of strongly interacting electrons (to which it can be mapped by Jordan-Wigner transformation). Understanding the Heisenberg model would therefore provide a guidance to physics of more realistic models of interacting electrons like e.g. the Hubbard model. 

Despite integrability of the Heisenberg model \cite{korepin}, calculating quantities beyond, for instance, the ground state energy or low-energy excitations is exceedingly difficult. Spin-spin correlation functions are particularly important quantities. Regarding equilibrium correlation functions in the Heisenberg model the following is known \cite{korepin}: (i) at infinite temperature there are no correlations, (ii) at finite temperature correlations asymptotically decay exponentially with distance, (iii) at zero temperature (ground state) correlations decay asymptotically as a power-law of the distance   in the gapless phase for 
$|\Delta|<1$, while they decay exponentially in the gapped phase of $|\Delta|>1$. For non-equilibrium stationary states almost nothing is known about the correlation functions, some preliminary hints can be found in~\cite{NDC}.  The aim of this work is to focus on long range correlations in NESS.

While the properties in close-to-equilibrium situation can be studied using a linear-response formalism, far from equilibrium one is forced to consider genuine nonequilibrium setting where the central system is coupled to reservoirs at different potentials, for instance, at the chain ends. We are going to take into account the Markovian reservoirs in an effective way by using the Lindblad equation. Time evolution of systems's density matrix is therefore governed by the Lindblad master equation~\cite{master}
\begin{equation}
\frac{{\rm d}}{{\rm d}t}{\rho}=\ii [ \rho,H ]+ {\cal L}^{\rm bath}(\rho).
\label{eq:Lin}
\end{equation}
The dissipator ${\cal L}^{\rm bath}={\cal L}^{\rm L}+{\cal L}^{\rm R}$ is a sum of two parts, one acting on the left end  and the other acting on the right end of the chain. Each one is written in terms of a set of Lindblad operators $L^{\lambda}_k$ as
${\cal L}^{\lambda}(\rho)=\sum_k \left( [ L_k^{\lambda} \rho,L_k^{{\lambda}\dagger} ]+[ L_k^{\lambda},\rho L_k^{{\lambda}\dagger} ] \right)$, $\lambda={\rm L},{\rm R}$.
Most of the time we shall use the {\em one-spin} bath for which Lindblad operators act only on a single spin. Two Lindblad operators at each end are
$L^{\rm L}_{1,2}=\sqrt{\gamma\,\, \Gamma^\lambda_{\pm}}\sigma_1^{\pm}$,
$L^{\rm R}_{1,2}=\sqrt{\gamma\,\, \Gamma^\lambda_{\pm}}\sigma_n^{\pm}$,
$\Gamma^\lambda_\pm=\sqrt{\frac{1\pm\tanh{\mu_{\lambda}}}{1\mp\tanh{\mu_{\lambda}}}}$,
 where $\sigma_k^\pm=\frac{1}{2}(\sigma^{\rm x}_k \pm \ii\, \sigma^{\rm y}_k)$. In numerical simulations we always consider strong coupling $\gamma=1$.
These operators are such the the stationary state of the (say, left) bath spin only, ${\cal L}^{\rm L}(\rho_1)=0$, is $\rho_1\propto \exp{(\mu_{\rm L} \sigma^{\rm z}_1)}$ with polarization $\ave{\sigma_1^{\rm z}}=\tanh{\mu_{\rm L}}$. One can argue that such a situation formally corresponds to an infinite temperature with fixed bath magnetizations. Later we shall also consider simple {\em two-spin} baths where two border spins are coupled to a bath in order to simulate the effect of finite temperature.
The choice of the bath is qualitatively inessential for the  results shown.

To find NESS (fixed point of the Liouvillean) we simulate the Lindblad equation (\ref{eq:Lin}) using time-dependent DMRG (tDMRG)~\cite{Daley} with matrix product operator (MPO) ansatz. This enables us to study systems much larger than it would be possible with brute-force diagonalization. After sufficiently long time, the density matrix $\rho(t)$ converges to a time-independent NESS. Once NESS density matrix $\rho_{\rm NESS}$ is obtained in terms of MPO, various expectations can be efficiently evaluated.~\footnote{
In order to make sure that the truncation errors in tDMRG algorithm are under control, we carefully checked our tDMRG results against the result of an exact calculation for small systems up to $n\le 10$.
In each tDMRG run we also ensured that the results converged with respect to increasing MPO dimension $D$ (in some cases, we went up to $D=150$) and time of simulation.}

\begin{figure}[h!]
\centerline{\includegraphics[width=0.46\textwidth]{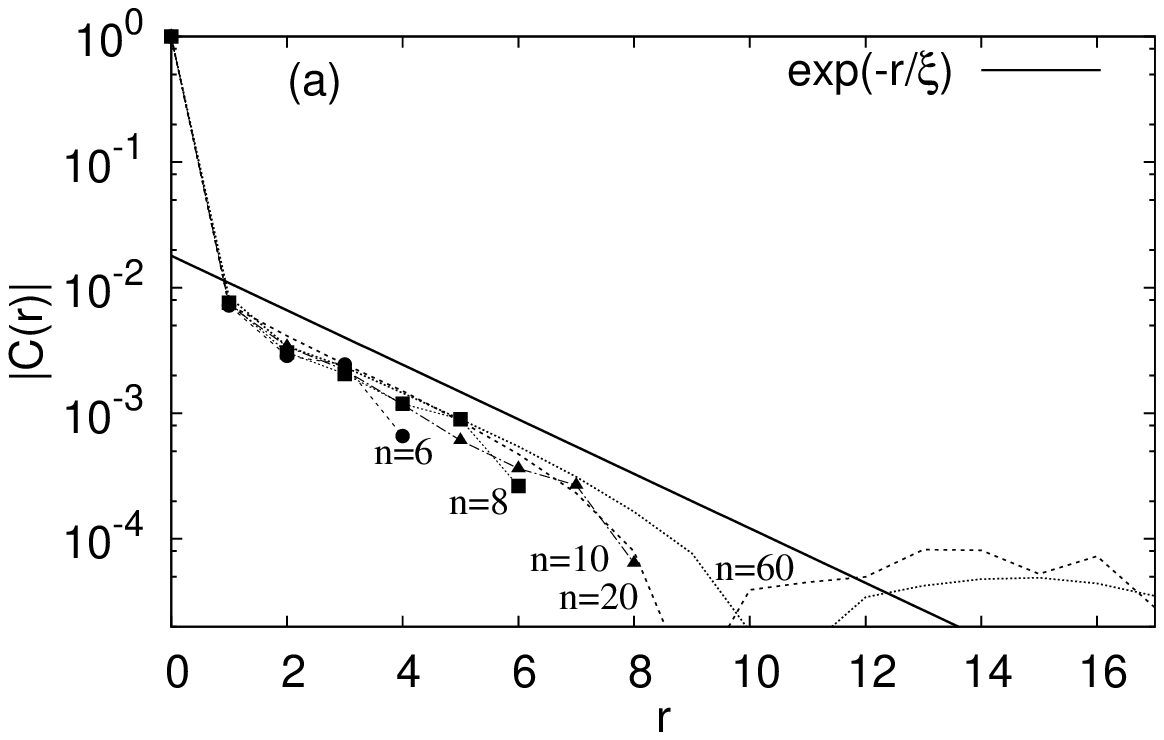}}
\vspace{-2mm}
\centerline{\includegraphics[width=0.46\textwidth]{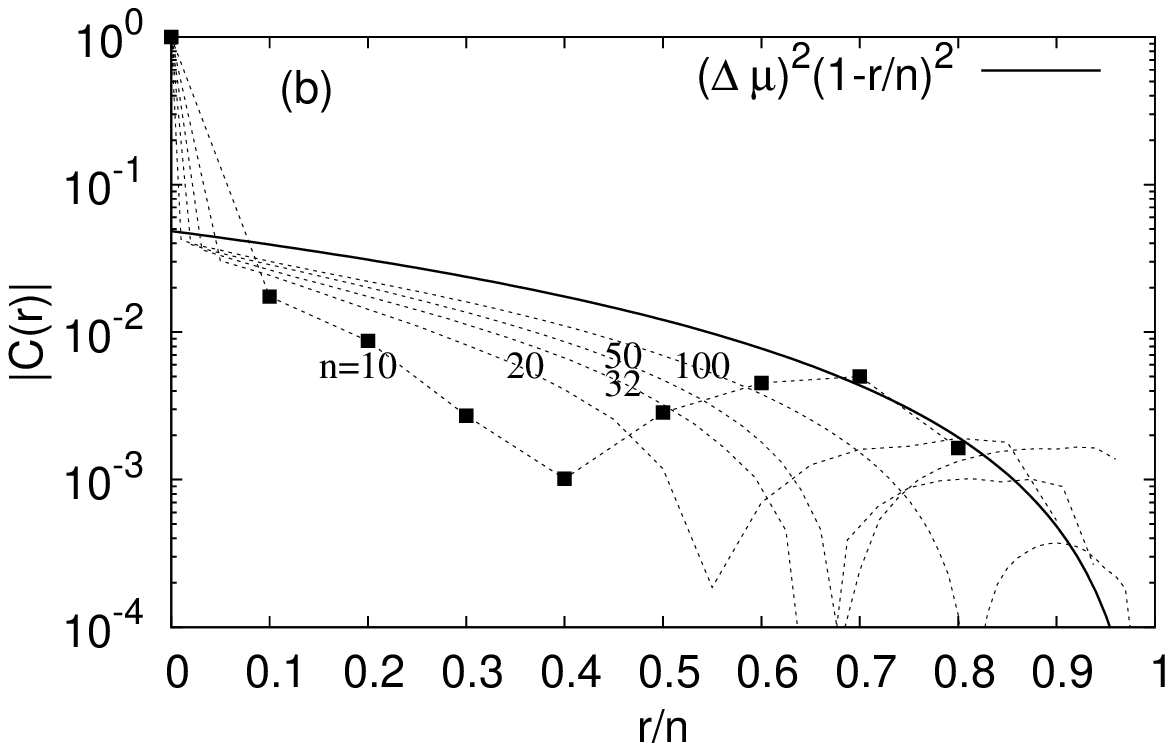}}
\caption{Spin-spin correlation function $C(r)$ in open XXZ model at $\Delta=0.5$ (a) and $\Delta=1.5$ (b), both for $\mu_{\rm L,R}=\pm 0.22$. Symbols are from exact calculation, dashed curves from tDMRG,
while full curves suggest $n\to\infty$ asymptotics.
}
\label{fig:d05}
\end{figure}

We now turn to the study of the main object of the present work, the 2-point spin-spin correlation function
\begin{equation}
C(i,j)= \tr \rho_{\rm NESS} \sigma_i^{\rm z} \sigma_j^{\rm z}-(\tr \rho_{\rm NESS}\sigma_i^{\rm z})(\tr\rho_{\rm NESS}\sigma_j^{\rm z}).
\label{eq:C}
\end{equation}
$C(i,j)$ is usually found to depend on the relative distance $r=|i-j|$ much stronger than on the perpendicular coordinate, thus we often show a 
one-dimensional cross-section of $C(i,j)$ along the skew-diagonal, that is $C(r):=C(k,k+r)$, where  $k=\lfloor \frac{n+1}{2}\rfloor -\lfloor \frac{r}{2}\rfloor$. Sometimes, when larger fluctuations among the
values of $C(i,j)$ are noticed, we average the absolute value of the correlation over all pairs of points
with a constant distance $\ave{|C(r)|} := \ave{|C(k,k+r)|}_k$, and in order to characterize LRO, as {\em order parameter} we compute the {\em residual correlator} as an aditional average over $r$ between $n/4$ and $3n/4$, 
$\ave{|C|} := \ave{|C(k,k+r)|}_{k}^{n/4\le r\le 3n/4}$.

It turns out that the spin-spin correlation function in NESS depends  strongly on the anisotropy $\Delta$.  
For example, for $\Delta=0.5$ the correlation function (its cross section $C(r)$ is shown in Fig.~\ref{fig:d05}a) shows exponential decay, with asymptotic scaling $C(r) \sim {\rm e}^{-r/\xi}$, with $\xi = 2.0$.  We can see that the correlation decay rate $\xi$ is independent of the chain size. Therefore, correlations in NESS are short-ranged. However, after increasing the anisotropy the situation becomes dramatically different, for example, for $\Delta=1.5$ the correlator $C(r)$ clearly exhibits LRO as shown in Fig.~\ref{fig:d05}b.
With increasing the chain size $n$ the correlation function converges to the limiting form for which we can observe several things: (i) spatial dependence is a function of the scaled variable $r/n$ {\em only}, for large $n$, meaning that there is LRO in NESS. 
\begin{figure}[h!]
\centerline{\includegraphics[width=0.46\textwidth]{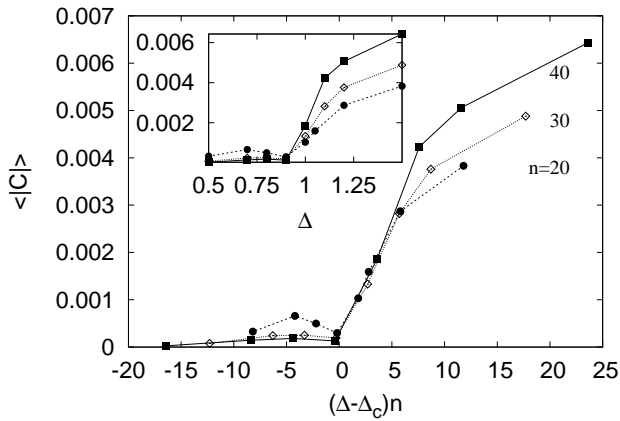}}
\caption{Residual correlator $\ave{|C|}$ in open XXZ model against the scaling variable $(\Delta-\Delta_{\rm c})n$ with $\Delta_{\rm c}=0.91$, for $n=20$ (circles), $30$ (diamonds), $40$ (squares). The inset shows unscaled data.
Here  $\mu_{\rm L,R}=\pm 0.22$ whereas practically the same scaling of $\ave{|C|}/(\Delta\mu)^2$ was obtained also for $\mu_{\rm L,R}=\pm 0.05$.
}
\label{fig:prehod}
\end{figure}
(ii) the size of the correlations scales with the driving as $\sim (\Delta \mu)^2=(\mu_{\rm L}-\mu_{\rm R})^2$ and is independent of $n$. Furthermore, correlations are positive (the point where they become negative moves towards $r/n \to 1$ as $n$ is increased). The limiting form of the correlation function is perhaps close to $C(i=n x,j= n y) \sim (\Delta \mu)^2 x(1-y)$. 
Very similar form of the 2-point correlation function, however with a size dependent prefactor ($1/n$), has been found in certain non-interacting, dissipative (noisy) or even classical models: in a ballistic exactly-solvable XY model~\cite{XY}, numerically observed in the diffusive Heisenberg model with dephasing~\cite{njp} and analytically explained in a diffusive XX model with dephasing~\cite{tobe}, but also in some classical lattice gasses like simple exclusion processes \cite{derrida} or its quantum analogs~\cite{temme}. The fact that long range correlator {\em here} does not decay with $n$ perhaps hints at truly {\em quantum coherent character} of these correlations.

From Fig.\ref{fig:d05} we can conclude that somewhere between $\Delta=0.5$ and $\Delta=1.5$ the range of correlations must change from short to long range, i.e., there is possibility of a nonequilibrium phase transition. To find if this is the case and where is the transition point we study the residual correlator
$\ave{|C|}$ against parameter $\Delta$ for several system sizes. Remarkably, numerics (shown in Fig.~\ref{fig:prehod}) indicates the
scaling $\ave{|C|} \sim f( (\Delta-\Delta_{\rm c})n^\nu)$, where $f$ is some universal scaling function and $\nu \approx 1.0$,
suggesting that the transition becomes {\rm non-continuous} in the thermodynamic limit.
Results seem to be accurate enough to support the claim that the observed critical anisotropy $\Delta_{\rm c} \approx 0.91$ is distinct from the zero temperature critical anisotropy $\Delta^*=1$ where the ground state gap opens.

\begin{figure}[t!]
\centerline{\includegraphics[width=0.45\textwidth]{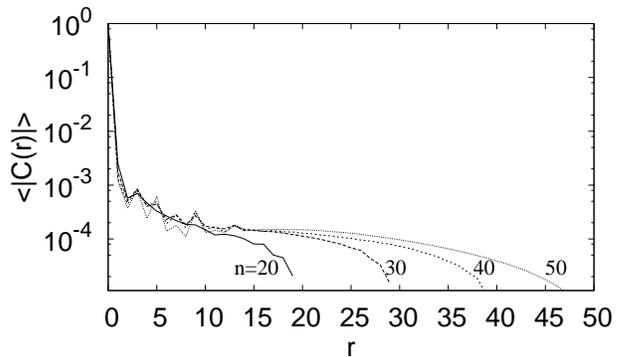}}
\caption{Smoothed spin-spin correlation function $\ave{|C(r)|}$  in the non-integrable deformed XXZ model with $\Delta=0.5$ and staggered transvere magnetic field $h=-0.5$, $\mu_{\rm L, R}=\pm 0.1$.}
\label{fig:stagg}
\end{figure}

One may  argue that LRO observed above might be due to integrability of the XXZ model. In order to check the effect of (non)integrability we
add a staggered magnetic field to the Hamiltonian (\ref{eq:H}), namely $H' = H + \sum_{j=1}^n h_j \sigma^{\rm z}_j$ with $h_{2j} = h$, $h_{2j+1} = 0$,
for $h=-0.5$, which is known to bring the model in the regime of quantum chaos \cite{NDC}. Using the same driving as before, we find that NESS again exhibits long range correlations. In order to somewhat isolate the effect of integrability breaking alone, we take $\Delta=0.5$, which in the integrable limit exhibited short range correlations. In order to smoothen the oscillations in the correlation function due to staggered field we plot in Fig.~\ref{fig:stagg} the average correlator $\ave{|C(r)|}$ and clearly observe LRO which is again consistent with a constant residual correlator $\ave{|C|} = {\cal O}(n^0)$.

\begin{figure}[t!]
\centerline{\includegraphics[width=0.46\textwidth]{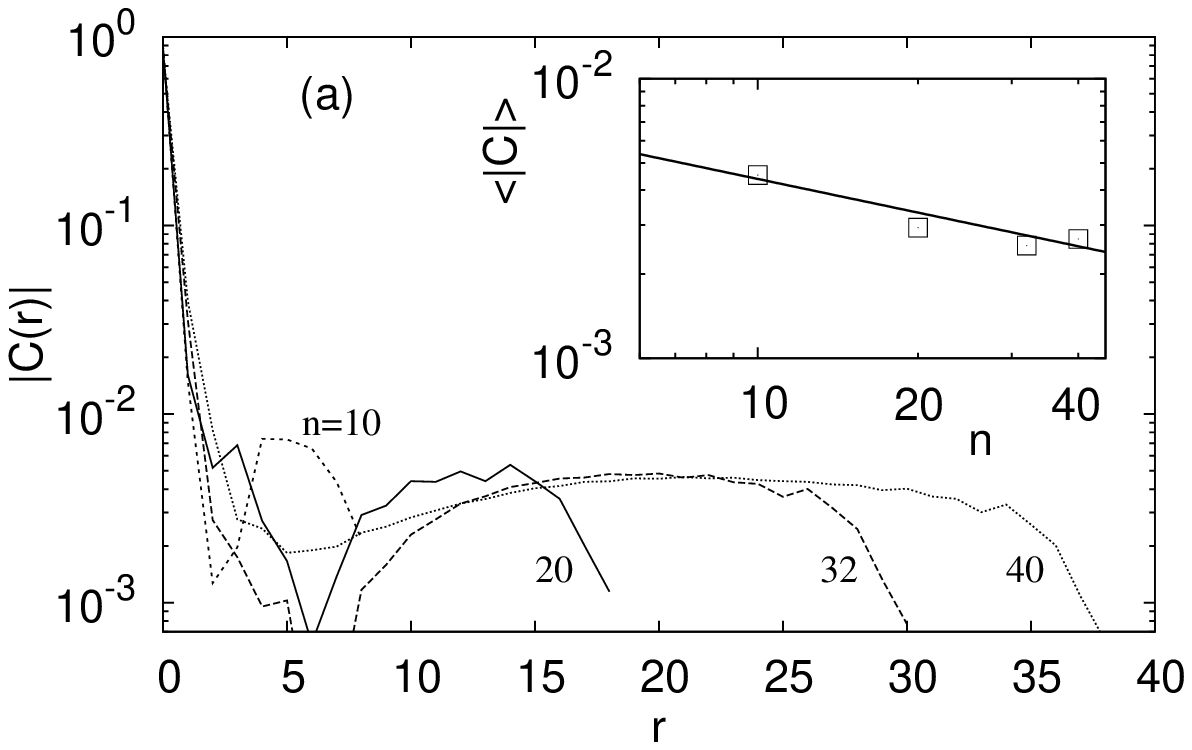}}
\vspace{-2mm}
\centerline{\includegraphics[width=0.46\textwidth]{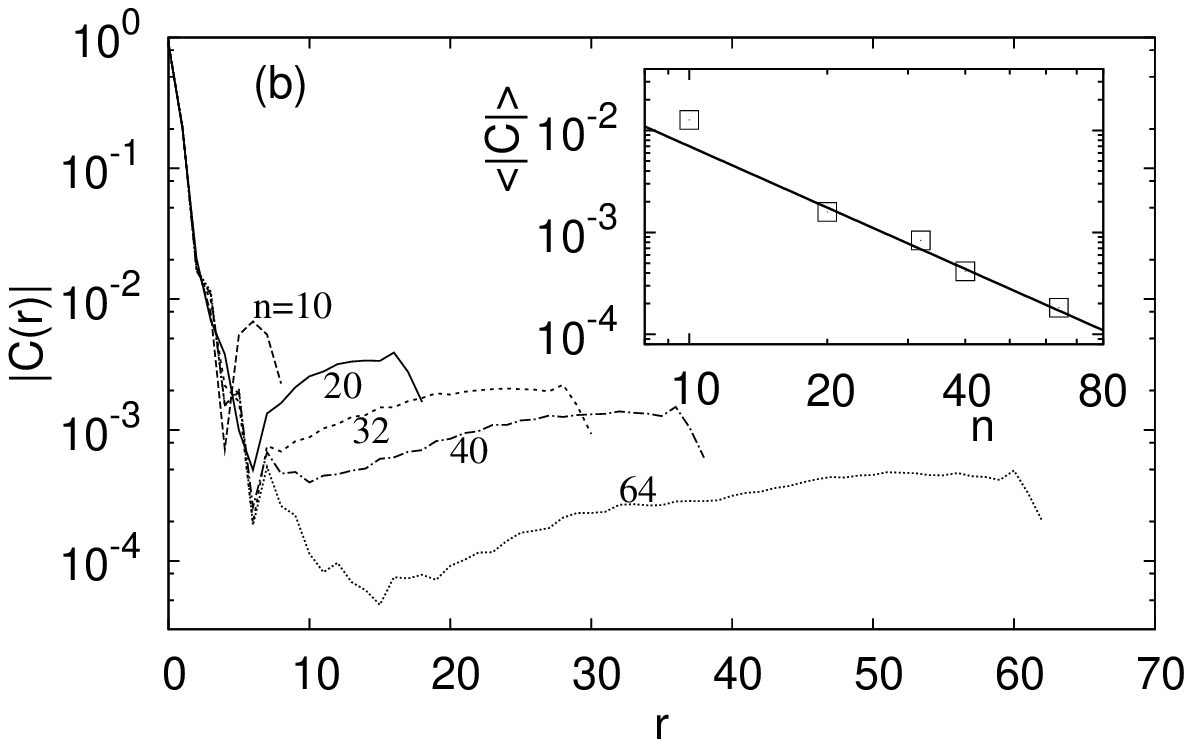}}
\caption{Finite temperature correlation function $C(r)$ for open XXZ model at $\Delta=1.5$. Frame (a) is for $T_{\rm meas} \approx 33$ while (b) is for $T_{\rm meas} \approx 5.8$ (both measured in the middle of the chain). At short distances ($r < 10$ in (b)) we have an exponential decay of $C(r)$ with the size-independent decay rate. For large $r$, we find the residual correlator scaling as $\ave{|C|} \sim (\Delta \mu)^2/n^\alpha$ (insets), with $\alpha \approx 0.4$ in (a) and $\alpha\approx 2$ in (b).}
\label{fig:finiteT}
\end{figure}

All results presented so far have been for symmetric driving, $\mu_{\rm R}=-\mu_{\rm L}$, which results in NESS having average magnetization zero and the average energy density also zero. This can be interpreted as NESS at an infinite average temperature, although care must be taken when discussing the temperature in an integrable open system because thermalization is absent for un-adjusted reservoirs due to lack of ergodicity~\cite{thermalization}. 
Thus the structure of the ground state very likely plays no role in the above NESS correlations. An interesting question is how do the correlations in NESS change as we lower the effective temperature? Also, correlations could depend on the average magnetization, as does for instance the spin transport in a linear response regime. In the zero magnetization sector there appears to be a transition from ballistic spin transport for $|\Delta| < 1$ to a diffusive transport in the gapped regime $|\Delta| > 1$. On the other hand, for nonzero magnetization spin transport is always ballistic as exemplified by a nonzero Drude weight~\cite{Zotos:97}. To study correlation function in NESS at a nonzero energy density (i.e., at finite ``temperature'') we used ${\cal L}^{\rm bath}$ coupling to two spins at the boundary or a one-spin bath with non-symmetric driving. Similar results are obtained in both cases and we are going to show only the ones for a two-spin bath as the relaxation
  to NESS there is faster. Lindblad operators on two boundary spins are chosen in such way that the invariant state of these two spins would be a grand-canonical state at some target temperature and magnetization in the absence of the bulk Hamiltonian~\cite{JSTAT,njp}. 
Effective temperature field of NESS is then prescribed by equating the measured local magnetization and energy density in the bulk to the grand-canonical one at some ``measured'' temperature $T_{\rm meas}$ and potential $\mu_{\rm meas}$ through canonical density matrix $\rho \sim \exp{\left[ -(H-\mu S^{\rm z})/T\right]}$, see~\cite{thermalization} for details.
The results for correlation function at two different measured temperatures in the center of the chain are shown in Fig.~\ref{fig:finiteT} for the gapped XXZ model $\Delta=1.5$ at zero staggered field $h=0$.
These two results for finite ``temperature'' show that the decay of the correlation function has two regimes. For small distances $r$, we have a short-range exponential decay, with the decay rate depending only on the temperature and not the chain length. This can be interpreted as a finite-temperature {\em equilibrium} part of the correlation function. On the other hand, at larger distances long-range correlations still persist, 
but scaling algebraically with the system size as $\ave{|C|} \sim (\Delta \mu)^2/n^\alpha$, with some nonzero power $\alpha$. This second regime can be interpreted as being genuinely {\em nonequilibrium}. Noting that at an infinite temperature, Fig.~\ref{fig:d05}b, we have $\alpha=0$ as discussed before, while at lower temperatures, Fig.~\ref{fig:finiteT}, $\alpha$ increases. We conjecture that $\alpha$ goes from $0$ at an infinite temperature to $\alpha \to \infty$ at zero temperature. Increasing exponent $\alpha$ with decreasing temperature might be in line with the exponential decay of correlations for a gapped ground state in equilibrium. Note however, that in the equal-bath (equilibrium-like) situation, $\Delta \mu \to 0$, all correlations in NESS decay exponentially.


In conclusion, we have presented strong numerical support for the existence of long range correlations in boundary driven non-equilibrium steady states of locally interacting quantum spin chains. 
We have shown the existence of quantum phase transition in an open XXZ spin $1/2$ chain with spontaneous emergence of long range correlations with increasing the anisotropy, and/or breaking the integrability of the model. These long range correlations are particularly strong, as they - evaluated at the fixed scaled distance $r/n$ - do not depend in the system size $n$, and are thus fundamentally different from the `hydrodynamic' long range correlations observed for instance in certain classical far from equilibrium stochastic processes \cite{derrida}.
This phenomenon could have far reaching consequences for far from equilibrium physics in low dimensional quantum many-body systems, such as
complete absence of local equilibrium and non-existence of local thermodynamic potentials.
The work is supported by grants P1-0044, J1-2208 of Slovenian Research agency.

\end{document}